# COEXISTENCE OF SUPERCONDUCTIVITY AND MAGNETISM IN $R_{2-x}Ce_xRuSr_2Cu_2O_{10-d}$ (R=Eu and Gd)

I. Felner

Racah Institute of Physics, The Hebrew University, Jerusalem, 91904, Israel.

$R_{2-x}Ce_xRuSr_2Cu_2O_{10-\delta}$(Ru-2122) is the first Cu-O based system in which superconductivity (SC) in the $CuO_2$ planes and *weak*-ferromagnetism (W-FM) in the Ru sublattice coexists. The hole doping in the $CuO_2$ planes, is controlled by appropriate variation of the Ce concentration and/or increasing the oxygen concentration. In $Eu_{2-x}Ce_xRuSr_2Cu_2O_{10}$, SC occurs for Ce contents of 0.4-0.8, with the highest $T_C$=35 K for Ce=0.6. The as-prepared non-SC $EuCeRuSr_2Cu_2O_{10}$ sample exhibits magnetic irreversibility below $T_{irr}$=125 K and long range o *anti*-ferromagnetic (AFM) order at $T_M$ =165 K. The saturation moment at 5 K is $M_{sat}$=0.89 $\mu_B$ /Ru. Annealing under oxygen pressures, does not affect these parameters, whereas depletion of oxygen shifts both $T_{irr}$ and $T_M$ up to 169 and 215 K respectively. $T_M$, $T_{irr}$ and $M_{sat}$ decrease with x, and the Ce dependent magnetic-SC phase diagram is presented. A simple model for the SC and the long-range W-FM states is proposed. We argue that: (i) the system becomes AFM ordered at $T_M$; (b) at $T_{irr}$ <  $T_M$, W-FM is induced by the canting of the Ru moments, and (c), at lower temperatures the appropriate samples become SC at $T_C$. The magnetic features are not affected by the SC state, and the two states coexist.

## Introduction

The interplay of magnetism and superconductivity (SC) is a fundamental problem in condensed - matter physics and it has been studied experimentally and theoretically for almost four decades. These two cooperative phenomena are mutually antagonistic. SC is associated with the pairing of electron states related to time reversal, while in the magnetic states the time-reversal symmetry is lost and therefore there is a strong competition with SC. Indeed, in conventional superconductors, local magnetic moments break up the spin-singlet Cooper pairs and hence strongly suppress SC, an effect known as pair-breaking. Therefore, a level of magnetic impurity of only 1 %, can result in a complete loss of SC. In a limited class of inter-metallic systems, SC occurs even though magnetic ions with a local moment occupy all of one specific crystallographic site, which is well isolated and de-coupled from the conduction path. The study of this class of magnetic-superconductors was initiated by the discovery of $RRh_4B_4$ and $RMo_6S_8$ compounds (R=rare-earth), and has been recently revitalized by the discovery of the $RNi_2B_2C$ system. In all three systems, both SC and antiferromagnetically (AFM) ordered states coexist. The onset of SC takes place at Tc ~ 2-15 K, while AFM order appears at lower temperatures (except for $DyNi_2B_2C$); thus, the ratio $T_N/T_C$ is ~ 0.1-0.5. Many of the high $T_C$ superconducting materials (HTSC) contain magnetic R constituentsand are AFM ordered at low temperatures, e.g. in $GdBa_2Cu_3O_7$ ($T_C$ = 92 K), and $T_N$ (Gd)=2.2 K. The R sublattice is electronically isolated from the Cu-O planes, and has no adverse effect upon the superconducting state.

Coexistence of weak-ferromagnetism (W-FM) and SC was discovered a few years ago in $R_{2-x}Ce_x$**Ru**$Sr_2Cu_2O_{10}$ (R=Eu and Gd, Ru-2122) layered cuprate systems [1-2], and more recently [3] in $GdSr_2RuCu_2O_8$ (Ru-1212). The SC charge carriers originate from the $CuO_2$ planes and the W-FM state is confined to the Ru layers. In both systems, the magnetic order does not vanish when SC sets in at $T_C$, and remains unchanged and coexists with the SC state to within a region of (at least) 10 micrometers. The $Eu_{2-x}Ce_x$**Ru**$Sr_2Cu_2O_{10}$ display a magnetic transition at $T_M$= 125-180 K and bulk SC below $T_C$ = 32-50 K, depending on oxygen concentration and sample preparation [1]. Here $T_M/Tc$ ~4, a trend which is contrary to that observed in the inter-metallic systems. The hole doping of the Cu-O planes, which results in metallic behavior and SC, can be optimized with appropriate variation of the R/Ce ratio [4-5]. SC occurs for Ce contents of 0.4-0.8, and the highest $T_C$ was obtained for Ce=0.6. K. Specific heat studies show a sizeable typical jump at $T_C$ and the magnitude of the $\Delta C/T$ (0.08 mJ/gK$^2$) indicates clearly the presence of bulk SC [6]. The specific heat anomaly is independent of the applied magnetic field. SC survives because the Ru moments probably align in the basal planes, which are practically de-coupled from the $CuO_2$ planes, so that there is no pair breaking. Scanning tunneling spectroscopy[1], muon-spin rotation [7], Raman [8] and magneto-optic experiments have demonstrated that both states coexist within the same crystalline grain. The isostructural compounds $R_{2-x}Ce_xMSr_2Cu_2O_{10}$ with M= Nb and Ta, are also SC below $T_C$~28-30 K, but do not show long-range magnetic order [9].

In the Ru-2122 materials, the W- FM state, as well as irreversibility phenomena, arise as a result of an antisymmetric exchange coupling of the Dzyaloshinsky-Moriya (DM) type [10] between neighboring Ru moments, induced by a local distortion that breaks the tetragonal symmetry of the $RuO_6$ octahedra.

The most remarkable magnetic properties of the SC Ru-2122 materials are: (a) A negative magnetic moment in the ZFC branches measured at low applied fields ($H_{ext}$), (b) ferromagnetic-like hysteresis loops and strong enhancement of the coercive field *in the SC state at T< $T_C$*; and (c) the presence of the so-called spontaneous vortex phase (SVP), which permits magnetic vortices to be present in equilibrium without an external magnetic field [11]. The vortices in the SC planes are caused by the internal field $B_{int}= 4\pi M$ (higher than $H_{c1}$) of a few hundreds of G of the FM Ru sublattice. (d) Also at relatively low H (~1-5 Oe), no diamagnetic signal, in the FC branch (the Meissner state (MS)- the conventional signature of a bulk SC), has been observed. The absence of the MS, may be a result of the SVP, and/or the high Ru moment induced by the external field at $T_M$, which masks this SC signature. On the other hand, when Ru is partially replaced by Nb, the small positive contribution of the W-FM Ru sublattice decreases the internal field and the MS is readily observed [12].

This paper covers the physical properties of the Ru-2122 system only and is organized as follows: In the next section, basic concepts of the crystal structure and of the Ru valence state are described. This section is intended for the non-specialist: those already working in this field can skip it. In sections II and III which are the central content of this paper, an assessment of the most reliable experimental results of the superconducting and magnetic states are described. Conclusions are summarized in Section IV.

## (I) BASIC CONCEPTS
**a. Crystal structure and X-rays data**

The tetragonal Ru-2122 structure (space group I4/mmm) evolves from the $RBa_2Cu_3O_7$ structure by inserting a fluorite type $R_{1.5}Ce_{0.5}O_2$ layer instead of the R layer in $RBa_2Cu_3O_7$, thus shifting alternate perovskite blocks by (a+b)/2 (Fig. 1). The magnetic Ru (as well as the M) ions reside in the Cu (1) site. Powder X-ray diffraction (XRD) measurements of $Eu_{2-x}Ce_xRuSr_2Cu_2O_{10-\delta}$ confirmed the purity of the compounds (~97%) and indicate, within the instrumental accuracy, that all samples have the same lattice parameters a=3.846(1) Å and c=28.72(1) Å. Due to the similarity of the ionic radii of $Eu^{3+}$ (0.94 Å ) and $Ce^{4+}$(0.87Å ), the lattice parameters of Ru-2122 materials are independent of x (and δ). The diffraction scans of $Eu_{1.5}Ce_{0.5}RuSr_2Cu_2O_{10-\delta}$ at various temperatures indicate clearly that all peaks and their relative intensities remain unaltered down to 10 K [13]. The diffraction scans do not show any anomalies around $T_C$ and $T_M$. The a and c lattice parameters are linear down to 100 K, and the typical flattening of the thermal expansion behavior is observed at low temperatures. The detailed crystal structure and the atomic positions in Ru-2122 were studied by synchrotron X-ray diffraction and neutron diffraction [14] experiments, which show that the $RuO_6$ octahedra are rotated ~ 14° around the c-axis and that this rotation is essentially the same for x=1 and x=0.6, as well as for Ru-1212. There is no evidence for super-cell peaks in the Ru-2122 samples.

**b. The *pentavalent* state Ru**

The valence of Eu, Ce and Ru ions in $Eu_{1.5}Ce_{0.5}RuSr_2Cu_2O_{10-\delta}$ have been studied by Mossbauer spectroscopy (MS) and X-ray-absorption spectroscopy (XAS) techniques [15]. MS performed on $^{151}$Eu show a single narrow line with an isomer shift =0.69(2) mm/s and a quadrupole splitting of 1.84 mm/s, indicating that the Eu ions are trivalent with a nonmagnetic J=0 ground state. This is in agreement with XAS taken at $L_{III}$ edges of Eu and Ce which shows that Eu is trivalent and Ce is tetravalent. The local electronic structure in several Ru based compounds was studied by XAS at the K edge of Ru, and the results obtained at room temperature are shown in Fig. 2. Since the valence of $Gd^{3+}$, $Sr^{2+}$ and $O^{2-}$ are conclusive, a straightforward valence counting for $GdSr_2RuO_6$ and $SrRuO_3$ yields Ru, as $Ru^{+5}$ and $Ru^{+4}$ ions respectively. The similarity between the XAS spectra of Ru-2122 and $GdSr_2RuO_6$ indicates clearly, that in Ru-2212 the Ru ions are in a *pentavalent* state. It is apparent that SC in the M-2122 system exists only for pentavalent M ions such as Nb, Ta and Ru. Recent $L_{III}$-edge XANES spectra on $Gd_{2-x}Ce_x$**Ru**$Sr_2Cu_2O_{10}$ (x=1 and 0.5) show that Ru is pentavalent irrespective of the Ce concentration[**]. This implies that there is no charge transfer to the RuO layer with increasing Ce concentration.

**c. Experimental Details**

Samples of $R_{2-x}Ce_xRuSr_2Cu_2O_{10-\delta}$ were prepared by a solid state reaction technique as described elsewhere [1,4]. Parts of the as-prepared (asp) samples were re-heated for 24 h at 800° -825° C under various pure oxygen pressures up to 150 atm. and will be identified according to the applied pressure. Determination of the absolute oxygen content in the asp materials and in the high oxygen pressure annealed samples, is difficult because $CeO_2$ is not completely reducible to a stoichiometric oxide when heated to high temperatures. Thermo-gravimetric measurements of $Eu_{1.5}Ce_{0.5}RuSr_2Cu_2O_{10-\delta}$ show that the materials are stable up to 600°C and no oxygen weight loss is detected. Above this temperature a small weight decrease begins. Our analysis indicates that the sample annealed at 150 oxygen atm. contains ~4 at % more oxygen than the asp one. ZFC and FC dc magnetic measurements in the range of 5-300 K were performed in a commercial (Quantum Design) super-conducting quantum interference device (SQUID) magnetometer. The resistance was measured by a standard four contact probe and the ac susceptibility was measured by a home-made probe with excitation frequency and amplitude of 733 Hz and 30 mOe respectively, both inserted in the SQUID magnetometer.

## (II) SUPERCONDUCTIVITY IN $R_{2-x}Ce_xRuSr_2Cu_2O_{10-\delta}$ RESULTS AND MODEL

### The Effect of Ce, Oxygen and pressure on $T_C$

The temperature dependence of the normalized real ac susceptibility curves (at $H_{dc}=0$) of $Eu_{2-x}Ce_xRuSr_2Cu_2O_{10-\delta}$ are presented in Fig. 3. It is readily observed that the x=1 and 0.9 samples are not SC and that SC occurs for Ce contents of x=0.8-0.4. Fig 4. exhibits the onset of SC deduced from these ac plots which exhibit a bell shape behavior with a peak at 35 K for Ce=0.6. Similar values are obtained by resistivity measurements. A similar bell shape behavior was observed in the non magnetic $Eu_{2-x}Ce_xNbSr_2Cu_2O_{10-\delta}$ (x=0.4-1) system, which serve as our reference materials. This may suggest that $T_C$ is not being significantly affected by the magnetic state of the Ru-O layers discussed below(e.g. by possible magnetic pair-breaking).

The external pressure dependence of $T_C$ in $Eu_{1.5}Ce_{0.5}RuSr_2Cu_2O_{10}$ is linear up to 2.5 GPa. (Fig.5). The pressure coefficient of 4.7 K/GPa obtained [17], is two times larger than that of Ru-1212.

The temperature dependence of the normalized resistance R(T) for the (asp) $Eu_{1.5}Ce_{0.5}RuSr_2Cu_2O_{10-\delta}$ and 22 atm. samples (measured at H=0 ) is shown in Fig. 6. The onset of the SC transition for the asp ($T_C$ = 32 K) is shifted to 38 K. At high temperatures, a metallic behavior is observed, and for the asp sample, an applied field of 5 T smears the onset of SC and shifts it to 28 K. The SC transition for the asp sample is more easily seen in the derivative dR/dT plotted in the inset. At $T_C$ = 32 K the derivative rises rapidly and does not fall to zero until the percolation temperature around 19 K is obtained. This behavior is typical for under-doped HTSC materials, where inhomogeneity in oxygen concentration causes a distribution in the $T_C$ values. It also may be related to the granular character of the poly-crystalline materials discussed below. The dependence of $T_C$ on the applied oxygen pressure obtained from resistivity measurements is presented in Fig. 7, exhibiting a monotonic increase from 32 to 49 K.

**b. Physical properties of the SC state.**

Specific heat studies of 2122 (Fig. 8) show a sizeable typical jump at $T_C$ =34 K for $Gd_{1.4}Ce_{0.6}RuSr_2Cu_2O_{10}$ and the magnitude of the $\Delta C/T$ (0.08 mJ/gK$^2$) indicates clearly the presence of bulk SC [6]. The specific heat anomaly is independent of the applied magnetic field (up to 6 T). The absolute thermo-power S(T) for $Eu_{1.4}Ce_{0.6}RuSr_2Cu_2O_{10}$ ($T_C$ =45 K) annealed under an oxygen pressure of 54 atm., is positive over the entire temperature range (Fig. 9). The positive S(T) confirms that the carriers are holes (see below). The feature in the S(T) curve is a break in the slope at $T_C$ followed by a drop to zero at 25 K. This relative broad transition suggests that the oxygen in the sample may be inhomogeneous. Alternatively, this may be a characteristic of the coupled SC states identified as the spontaneous vortex phase discussed in details in Ref. 12, or to the granular nature of this material. The small bump at 140 K is associated with the magnetic transition of this material. The thermopower is quite linear with positive slope between $T_C$ and 100 K without any clear feature at $T_{irr}$.

**c. The reason for superconductivity in the $M^{5+}$-2122 (M=Ru, Nb and Ta) systems**

Given the variety of crystal structures and the chemical methods used to introduce holes into the $CuO_2$ layers, it is well accepted that a "generic' electronic phase diagram can be sketched for all compounds. Hole (or carrier) density (*p*) in the $CuO_2$ planes, or deviation of the formal Cu valence from $Cu^{2+}$, is a primary parameter which affects $T_C$ in most of the HTSC compounds. In the well-established phase diagram $La_{2-x}Sr_xCuO_4$, the insulating parent $La_2CuO_4$ is AFM ordered. The magnetic interactions are well described by a simple Heisenberg model, with a large exchange interaction (J= 1500 K) value. In $La_{2-x}Sr_xCuO_4$ the charge carrier concentration, can be varied by replacing $Sr^{2+}$ for $La^{3+}$. The variation of $T_C$ as a function of hole doping exhibits a bell shape behavior, with a peak for the optimally doped material (x=0.15).

We argue that the RCe**M**Sr$_2$Cu$_2$O$_{10}$ (R/Ce=1) samples serve as the parent stoichiometric insulator compounds (similar to $La_2CuO_4$). Since the valence of $R^{3+}$, $Ce^{4+}$, $M^{5+}$ $Sr^{2+}$ $Cu^{2+}$ and $O^{2-}$ are conclusive, a straightforward valence counting yields a fixed oxygen concentration of 10. Hole doping of the Cu-O planes, which results in metallic behavior and SC, can be optimized with appropriate variation of the $R^{3+}/Ce^{4+}$ ratio (trivalent $R^{3+}$ ions are replaced for $Ce^{4+}$). SC occurs for Ce contents of 0.4-0.8 (Fig. 3-4) and the optimally doped sample is obtained for Ce=0.6. This picture is consistent with the positive S(T) curve shown in Fig. 9. Unlike the $La_{2-x}Sr_xCuO_4$ system, this substitution does not appear to significantly alter the hole carriers on the Cu-O planes. This is apparent in Fig. 4 which shows that the change of x from 0.8 to 0.6, results in a small increase in $T_C$. Indeed, if all the carriers were introduced into the Cu-O planes, then for the under-doped (x=0.8) to the optimally doped (x=0.6) samples, p should vary by 0.2 and result in a large shift in $T_C$, as observed in $La_{2-x}Sr_xCuO_4$ and in other HTSC materials. It is thus possible, that in all M-2122 compounds, the extra holes introduced by reducing the Ce content, are *partially* compensated for by depletion of oxygen [16] and in $R_{2-x}Ce_xMSr_2Cu_2O_{10-\delta}$ the oxygen deficiency (d) increases with $R^{3+}$. This partial hole doping mechanism is also supported by neutron diffraction on $Gd_{1.3}Ce_{0.7}RuSr_2Cu_2O_{10-\delta}$ from which one finds $\delta$=0.22, instead of 0.3 required by charge neutrality [14].

**d. The granular behavior of the SC transition**

Due to the weakly coupled granular superconductivity of the ceramic Ru-2122 samples, the intrinsic properties of the SC state are further complicated. The microstructure of Ru-2122 (x=0.5) exhibits well-defined grains with a typical size of a few *m*m and pronounced grain boundaries [18]. The transport transition to the SC state occurs via two intermediate stages [19]. The intra-grain sharp drop of the resistivity at $T_C$ occurs when the grains become superconducting. At this point the weak-links (WLNs) between the grains contribute non-zero resistance across the sample. At lower temperature ($T_P$) the WLNs become superconducting and the inter-grain transition is much more sensitive to the magnetic field. The broadening of the resistive transition was already observed in oxygen annealed Ru-2122 materials [1], and was related to inhomogeneity in the oxygen concentration which causes a distribution in WLNs and in the $T_C$ values.

Fig. 10. shows the temperature dependence of real ac susceptibility of asp $Eu_{1.4}Ce_{0.6}$**Ru**$Sr_2Cu2O10_{-\delta}$ at various amplitudes (up to $H_{ac}$ =3.6 Oe). It is readily observed that the broad transition for this granular superconductor occurs via two stages. The onset of superconductivity ($T_c$= 36 K) at which the grains become superconducting, is not severely affected by $H_{ac}$. However, the step-like transition at a much lower temperature ($T_P$ =27 K), which is due to weak-Josephson inter-grain coupling, is affected dramatically by the applied field. This is typical for a granular superconductor with weak inter-grain coupling. Generally speaking, below $T_P$ the susceptibility is governed mainly by the WLN properties and it is sample dependent. Various parameters affect the WLN properties such as: heat treatment, annealing conditions and the way the samples have been stored prior to measuring.

The transport critical current density ($J_C$) at 5 K, of $Gd_{1.4}Ce_{0.6}$**Ru**$Sr_2Cu_2O_{10-\delta}$, was measured by the I-V curves at various applied fields (Fig. 11). The onset of the resistivity is changed dramatically to lower currents with the applied field below 10 Oe, but is almost unchanged for $H_{ext}$ higher than 20 Oe (not shown). The deduced field dependence of $J_C$ is exhibited in Fig. 12, and the solid line is the fit of curve to: $J_C \propto H_{ext}^{-0.45}$. The value obtained at 5 K and H=0 (22 A/cm$^2$) is a few orders of magnitude lower than the measured $J_C$ of ceramic $YBa_2Cu_3O_7$ [20]. The intergrain $J_C$ of Ru-2122 can be evaluated from the remanent moment (at 5 K) of the SC Nb-1222 which serves as a reference material, assuming similar SC properties of the two iso-structural systems. The remanent field obtained for $Gd_{1.4}Ce_{0.6}$**Nb**$Sr_2Cu_2O_{10-\delta}$, is 17.5 Oe. In the Bean model $J_C=30*\Delta M/d$ (where d is a typical particle size parameter). Taking d as 0.3 cm (the largest dimension in the sample) we obtain $J_C\sim$ 1575 A/cm$^2$, a value which is much higher than those in Fig. 3. Therefore, it is possible that the low $J_C$ of Ru-1222 is related to its magnetic behavior and to the SVP [12]. More investigation is needed to clarify this point.

### (III) WEAK-FERROMAGNETISM IN $R_{2-x}Ce_xRuSr_2Cu_2O_{10-d}$ RESULTS AND MODEL

In contrast to the Ru-1212 system in which the antiferromagnetic nature of the Ru sublattice has been determined by neutron diffraction studies [21-22], the published data up to now have not include, any determination of the magnetic structure in Ru-2122. Since all the materials have basically the same complicated magnetic behavior, we shall start to describe the magnetic properties of non-superconducting $EuCeRuSr_2Cu_2O_{10}$ (x=1) material.

**a. The magnetic properties of $EuCeRuSr_2Cu_2O_{10}$**

ZFC and FC dc magnetic measurements (at 50 Oe) for the parent as-prepared (asp) $EuCeRuSr_2Cu_2O_{10}$ sample, are shown in Fig.13. The two curves merge at $T_{irr}$=125 K. Note the ferromagnetic-like shape of the FC branches. The magnetic transition- $T_M(Ru)$ is not at $T_{irr}$. The M/H(T) curves do not lend themselves to an easy determination of $T_M(Ru)$, and $T_M(Ru)$=165 K, was obtained directly from the temperature dependence of the saturation moment ($M_{sat}$), discussed below. [An alternative way to determine $T_M(Ru)$ is to cool the material from above $T_M$, under a small *negative* magnetic field (say -5 Oe), which aligns the Ru sublattice. At low temperatures, a small *positive* (5 Oe) is applied. Due to the high anisotropy, the Ru moments remain opposite to the field direction up to $T_M$. The measured negative M(T) curve becomes zero at $T_M$]. $T_{irr}$ is field dependent, and shifted to lower temperatures with the applied field. $T_{irr}$ =91, 64, 39, 28 and 14 K for H= 250, 500, 1000, 2000 and 3000 Oe respectively. For higher external fields the irreversibility is washed out, and both ZFC and FC curves collapse to a single ferromagnetic-like behavior.

Part of the asp $EuCeRuSr_2Cu_2O_{10}$ sample was re-heated for 24 h at 800° C under high oxygen pressure (60 atm.) and another part was quenched from 1050° C to room temperature, denoted as hop and quenched materials respectively. Similar M/H(T) curves were performed to the oxygen annealed (hop) sample, and Fig. **3** shows that both $T_{irr}$ and $T_M(Ru)$ remain unchanged. This is in contrast to $Eu_{1.5}Ce_{0.5}RuSr_2Cu_2O_{10-\delta}$ where annealing under oxygen pressure affects $T_{irr}$ and $T_M(Ru)$ and shift them to higher temperatures [4]. This indicates (see above), the oxygen concentration in asp $EuCeRuSr_2Cu_2O_{10}$ is fixed and does not change during the annealing process. On the other hand, during the quenching process, a small amount of oxygen is depleted and both $T_{irr}$ and $T_M(Ru)$ are shifted to 169 and 215 K respectively (Fig. 13). It is reminiscent of the magnetic phase-diagram of $YBa_2Cu_3O_z$, where the depletion of oxygen increases the magnetic transition of the $CuO_2$ planes.

M(H) measurements at various temperatures for the asp, hop and quenched samples have been carried out, and the results obtained for the asp sample are exhibited in Figs. 14-15. All M(H) curves below $T_M$, are strongly dependent on the field (up to 2-4 kOe), until a common slope is reached (Fig. 14 inset). M(H) can be described as: $M(H) = M_{sat} + \chi H$, where $M_{sat}$ corresponds to the W-FM contribution of the Ru sub lattice, and $\chi H$ is the linear paramagnetic contribution of Eu and Cu). The saturation moment obtained at 5 K is $M_{sat} = 0.89(1)\mu_B$. Similar M(H) curves have been measured at various temperatures and Fig.15 (inset) shows the decrease of $M_{sat}$ with temperature. $M_{sat}$ becomes zero at $T_M(Ru)=165(2)$. The same $M_{sat}$ and $T_M(Ru)$ values were obtained for the hop material. However, for the quenched material, the saturation moment at 5 K remains unchanged, but $T_M$ is shifted to 215 (2) K. Thus, only the magnetic transitions are sensitive to the oxygen concentration. The measured $M_{sat}= 0.89\mu_B$, is somewhat smaller than the fully saturated moment $1\mu_B$ expected for the low-spin state of $Ru^{5+}$, i.e. $g\mu_B S$ for g=2 and S=0.5. This means that a small canting on adjacent Ru spins occurs, and the saturation moments are not the full moments of the $Ru^{5+}$ ions. The exact nature of the local structural distortions causing DM exchange coupling in Ru-2122 is not presently known.

At low applied fields, the M(H) curve exhibits a typical ferromagnetic-like hysteresis loop (Fig. 16). The positive virgin curve at low fields, indicates clearly that SC is totally suppressed. Two other characteristic parameters of the hysteresis loops are marked, namely, the remanent moment, ($M_{rem} = 0.41\mu_B$/Ru) and the coercive field ($H_C$ = -190 Oe at 5 K). The large $M_{rem}/M_{sat}$ ratio (at 5 K) is consistent with ferromagnetic-like order in $EuCeRuSr_2Cu_2O_{10}$. The same $M_{rem}$ and $H_C$ values (at 5 K) were obtained for the hop and quenched materials. Fig. 14 shows the temperature dependence of $M_{rem}$ obtained at 5 K, which disappears at $T_{irr}$. M(H) curves measured at various temperatures yield the $M_{rem}(T)$ and $H_C(T)$ values which are plotted in Fig. 15 (inset). For both the asp and quenched samples, $M_{rem}(T)$ also disappear at $T_{irr}$, and $H_C(T)$ become zero around 80 K and 130 K respectively (Fig. 8). Generally speaking, the magnetic properties of Ru in $EuCeRuSr_2Cu_2O_{10}$ are all *enhanced,* as compared to the asp superconducting $Eu_{2-x}Ce_xRuSr_2Cu_2O_{10-\delta}$ samples with x<1, and the effect of Ce concentration on the magnetic parameters of the Ru sublattice will be discussed below.

In the paramagnetic range (above 165 K), the $\chi(T)$ curve measured at 10 kOe for asp EuCe**Ru**$Sr_2Cu_2O_{10}$, has the typical paramagnetic shape and adheres closely to the Curie-Weiss (CW) law: $\chi = \chi_0 + C/(T-\theta)$, where $\chi_0$ is the temperature independent part of $\chi$, C is the Curie constant, and $\theta$ is the CW temperature. The net paramagnetic Ru contribution to $\chi(T)$, was obtained by subtracting $\chi(T)$ of EuCe**Nb**$Sr_2Cu_2O_{10}$ (the reference material) from the measured data. This procedure yields: $\chi_0 = 0.0063$ and C=0.57(1) emu/mol Oe and $\theta = 146(1)$ K, which corresponds to an effective moment $P_{eff} = 2.13$ $\mu_B$. The positive $\theta$ obtained is in fair agreement with $T_M$ and indicates ferromagnetic interactions. But $P_{eff}$ is greater than the expected value of the low-spin state of $Ru^{5+}$ and S=0.5 ($P_{eff} = 1.73$ $\mu_B$).

**b. The Effect of Ce on the Magnetic Properties of $Eu_{2-x}Ce_xRuSr_2Cu_2O_{10-\delta}$**

All the $Eu_{2-x}Ce_xRuSr_2Cu_2O_{10-\delta}$ compounds reported here have been prepared simultaneously under the same conditions and our extensive magnetic study shows, that the magnetic behavior of all materials is quite similar to those described in Figs 13-17. For the sake of brevity, we display the in Fig. 17 only the data obtained for x=0.1 and x=0.5. It is readily observed, that for x=1, the magnetic properties due to the Ru are all *enhanced*, as compared to the x=0.5. The latter sample is SC and its ZFC branch starts from negative values. At 50 Oe. the diamagnetic signal due to high shielding fraction (SF) of the SC state dominates, and the net moment at low temperatures is negative. The weak ferromagnetic component of Ru and the high paramagnetic effective moment of $Eu^{3+}$ do not permit a quantitative determination of the SF from these curves. The inflection in the FC branch agrees well with $T_C$ determined from the ac curve (Figs.3,6).

The variation of $T_{irr}$ and $T_M$ as a function of Ce concentration in $Eu_{2-x}Ce_x\mathbf{Ru}Sr_2Cu_2O_{10-\delta}$, is summarized in Fig. 18. The enhancement of the magnetic properties is manifested by the monotonic rise of $T_{irr}$ and $T_M$ as x increases. $M_{sat}$ values at 5 K, increase gradually with x. It is apparent that this trend is not affected by the SC state which is induced for x=0.8. Due to the presence of a tiny amount of $SrRuO_3$ in the x=0.4 sample (not detectable by XRD), its $T_M$ was not determined. In the paramagnetic range, the C and $\theta$ parameters for all materials, are very close to those of the x=1 sample i.e. for x=0.5, C=0.58 emu/mol Oe, ($P_{eff}$=2.15 $\mu_B$) and $\theta$= 134(1) K), indicating similar net paramagnetic Ru contribution in all Ru-2122 compounds.

**c. Mossbauer effect of $^{57}Fe$ doped in $Gd_{1.4}Ce_{0.6}RuSr_2Cu_2O_{10-\delta}$**

Mossbauer effect studies (ME) on $^{57}Fe$ doped samples have been proved to be a powerful tool in the determination of the magnetic nature of the Fe site location. When the Ru ions become magnetically ordered, they produce an exchange field on the Fe ions residing in this site. The Fe nuclei experience a magnetic hyperfine field leading to a sextet in the observed ME spectra. As the temperature is raised, the magnetic splitting decreases and disappears at $T_M$. Fig.19 shows the real and imaginary ac susceptibility of $Gd_{1.4}Ce_{0.6}RuSr_2Cu_2O_{10-\delta}$, from which $T_{irr}$=92 and $T_M(Ru)$=175 K are deduced and Fig. 20 shows the ME studies on the $^{57}Fe$ doped material. The main effect to be seen in Fig. 20 is that the 4.2 and 180 K spectra consist of one site only. A least square fit to the spectrum at 180 K yields an isomer shift (IS) of 0.30(1) mm/s (relative to Fe metal) with a line width of 0.35 mm/s, and a quadrupole splitting ($\Delta_Q$ = ½eqQ) of 1.00(1) mm/s$^3$. This doublet is attributed to paramagnetic Fe ions in the Ru site [1]. At low temperatures, all spectra display magnetic hyperfine splitting, which is a clear evidence for long-range magnetic ordering. The fitting parameters of the single sextet obtained at 4.1 K are: IS= 0.40(1)mm/s, $H_{eff}(0)$=467(3) kOe and an effective quadrupole splitting value of $\Delta_{eff}$ = 1/2eQ$_{qeff}$ = -0.33(21) mm/s. Using the relation: $\Delta_{eff}= \Delta_Q/2(3\cos^2\theta-1)$, we obtained for the Ru site a hyperfine field orientation, $\theta$ = 72°, relative to the tetragonal symmetry c axis. As the temperature is raised, $H_{eff}$ decreases and disappears completely at $T_M(Ru)$= 175(5) K. $H_{eff}$ values obtained at 110, 130, 150 K and 160 K are: 399(3), 358(5), 312(2) and 279(5) kOe. Fig. 20 also shows that ME spectrum at 90 K (below $T_{irr}$) *is much different* than that at 110 K (above $T_{irr}$). In between these temperatures the canting spin reorientation occurs, well felt by the Fe probe.

**d. The Proposed Model of the Magnetic behavior of $Eu_{2-x}Ce_xRuSr_2Cu_2O_{10-d}$**

In contrast to the Ru-1212 system in which the antiferomagnetic nature of the Ru sublattice has been determined by neutron diffraction studies [21-22], the published data up to now have not include any determination of the magnetic structure in Ru-2122. Assuming a long-range order, the accumulated results are compatible with two alternative qualitative scenarios, both of which are used in understanding the qualitative features at low applied fields.

Going from high to low temperatures, the magnetic behavior is basically divided into 4 regions. (i) At elevated temperatures, the paramagnetic net Ru moment is well described by the CW law, and the extracted $P_{eff}$=2.15 $\mu_B$ and $\theta$=134-146 K values, practically do not alter with Ce content. (ii) At $T_M$ (depends on Ce content Fig. 18), the Ru sub lattice becomes AFM ordered. (iii) At $T_{irr} < T_M$, which is also Ce dependent and varies with the external field, a weak ferromagnetism is induced, which originates from canting of the Ru moments. $T_{irr}$ is defined as the merging point of the low field ZFC and FC branches, or alternatively, as the temperature in which the remanent moment disappears. This canting arises from the DM antisymmetric superexchange interaction, which by symmetry, follows from the fact that the $RuO_6$ octahedra tilt away from the crystallographic c axis. At high magnetic field (H>3000 Oe) the irreversibility is washed out and the M(T) curves exhibit a ferromagnetic-like behavior.(iv) At lower temperatures SC is induced and $T_C$ depends strongly on the $R^{3+}/Ce^{4+}$ (as hole carriers) and on oxygen concentrations. Below $T_C$, both SC and weak-ferromagnetic states coexist and the two states are practically decoupled. This model is supported by our Mossbauer studies (Fig. 20) and also from unpublished non-linear ac susceptibility measurements, which show non-linear signals up to $T_M$.

Our general picture is that in $Eu_{2-x}Ce_x\mathbf{Ru}Sr_2Cu_2O_{10-\delta}$ all compounds have a similar magnetic structure. We suggest two scenarios that could lead to the shift to higher values of the magnetic parameters (such as $T_M$, $T_{irr}$ and $M_{sat}$) with increasing Ce (Fig. 18). (a) The small difference between $Eu^{3+}$ and and $Ce^{4+}$ ionic radii, decreases the mean Ru-Ru distance, and as a result the magnetic exchange interactions become stronger with Ce, and (b) This trend arises from a change of the anti-symmetric exchange coupling (different oxygen concentration ?) of the DM type between the adjacent Ru moments, which causes the spins to cant out of their original direction to a larger (or smaller) angle, and as a result, a different component of the Ru moments forms the W-FM state.

Annealing the asp material under oxygen pressure, leads an enhancement of the $T_M$ values and changes other W-FM characteristic features. [1]. For the 75 atm oxygen annealed sample, both $T_{irr}$ and $T_M$ (Ru) are shifted to 137(2) and 168(2) K respectively. This effect, which was also observed in several rare-earth based inter-metallic hydrides, is probably an electronic effect. As described above, in addition to the change in the hole density of the Cu-O planes, there is a transfer of electrons from oxygen to the Ru 4d sub-bands, resulting in an increase of the exchange interactions between the Ru sublattice and hence to an increase in $T_M$ of the materials. An alternative way is to assume that this enhancement arises from an alternation of the anti-symmetric exchange coupling of the DM type between the adjacent Ru moments as discussed above. The exact nature of the local structure distortions causing the W-FM behavior in this system, as well as the extra oxygen location, are not presently known.

(2) Detailed analysis of the magnetization under various thermal-magnetic conditions suggests a phase separation of Ru-1222 into FM and AFM [23] nano-domain species inside the crystal grains. A minor part of the material becomes FM (at $T_M$), whereas the major part orders AFM and then becomes SC at $T_C$. In this scenario, the unusual SC state is well understood. Neutron diffraction measurements are required to precisely determine the nature of the magnetic order in the Ru-2122 system.

**e. The effect of hydrogen on the SC and magnetic behavior of Ru-2122**

Fig. 21 shows that in Ru-2122H$_{0.07}$, the effect of hydrogen is to suppress SC and to enhance the W-FM properties of the Ru sublattice (T$_M$ increased to 225 K) [24]. Two scenarios that could lead to this phenomenon are: (a) in addition to the change of $p$ in the CuO$_2$ planes, there is a transfer of electrons from hydrogen to the Ru 4d sub-bands, resulting in an increase in the Ru moments, and hence to enhance the magnetic parameters. Indeed, XAS data indicate, that in contrast to asp Ru-2122 (Fig. 2), the hydrogen loaded samples exhibit a dominant Ru$^{4+}$ valence [25]; (b) the enhancement arises from a change of the anti-symmetric exchange coupling of the DM type between the adjacent Ru moments, which causes the spins to cant out of their original direction to a larger angle. The picture which emerges from the STM measurements [1] is that hydrogen doping leads to phase separation. Even at very low doping (Ru-2122H$_{0.03}$), insulating regions start to form. As doping is increased, the density and size of the insulating regions increase, until they coalesce and the sample becomes globally insulating.

Fig. 22 presents the ZFC curves obtained for several hydrogen loaded samples. For the asp and the Ru-2122H$_{0.03}$ samples, the peak is around 80 K, and for the samples with H>0.14 at.% the peaks are shifted to about 160 K. For the intermediate hydrogen concentration (H=0.07) a superposition of both peaks is observed which leads to a somewhat flat curve.

The hydrogen atoms reside in interstitial sites, and their effect is reversible. Depletion of hydrogen leads to the original charge density and SC is restored. This is reflected in both the macroscopic magnetization studies and in the SC gap distribution extracted from STM result. T$_M$ drops back to 122 K and the peak in ZFC curve is shifted back to 80K [1]. Data for the regenerated sample are not presented here. Since hydrogen loading affects both (SC and W-FM) phenomena, we tend to believe that H atoms occupy interstitial sites, presumably inside the Sr-O planes.

(IV) **CONCLUSIONS**

We have shown that both SC and weak-ferromagnetism coexist in Ru-2122 and are an intrinsic property of this system. In contrast to other intermetallic magnetic-SC systems, the present materials exhibit magnetic order well above the SC transition (T$_M$/T$_C$ ~4). We attribute the magnetic order to the Ru sublattice, whereas SC is confined to the CuO2 planes. Both sites are practically decoupled from each other. Hole doping of the Cu-O planes, which results in SC, can be optimized with either (i) appropriate variation of the $Eu^{3+}/Ce^{4+}$ ratio and the optimally doped material is obtained for Ce=0.6, and (ii) by annealing under high oxygen pressures which leads to an increase in the oxygen concentration. The magnetic insulator parent EuCe**Ru**Sr$_2$Cu$_2$O$_{10}$ (x=1), is used to describe the magnetic behavior of the Ru-2122 system. For x=1 (δ=0), annealing under high oxygen pressure does not affect the magnetic properties, whereas $T_M$ is enhanced by oxygen depletion. The magnetic structure of all Eu$_{2-x}$Ce$_x$RuSr$_2$Cu$_2$O$_{10-\delta}$ materials studied is practically the same, but the magnetic parameters, such as $T_M$ and $M_{sat}$, increase with increasing oxygen (as well as hydrogen) content, but decrease with decreasing Ce content. In Ru-2122, the asp compound, is under-doped, and annealing under high oxygen pressure shifts $T_C$ to higher temperatures. On the other hand, the influence of hydrogen on Ru-2122 is reversible and not destructive, which means that hydrogen changes the hole density of the CuO$_2$ planes, either by increasing or decreasing the ideal effective charge of the planes. The exact nature of the Ru spins magnetic ordering is still debated and no conclusions has been reached as yet. Our preferred model is as follows. Two steps in the magnetic behavior are presented. At $T_M$ ranging from 122 K (for x=0.5) up to 225 K (for hydrogen loaded samples), all materials become AFM ordered. At $T_{irr}$ (depends on various parameters) a W-FM state is induced, originating from canting of the Ru moments. This canting arises from the DM anti-symmetric super-exchange interaction and follows from the fact that the RuO$_6$ octahedra tilt away from the crystallographic c axis. Neutron diffraction studies did not show any evidence for extra magnetic peaks [14]. It is possible that the ~Ru-2122 is an itinerant ferromagnetic system in which the magnetic moments are not localized. There are number of similarities between the magnetic and electronic behavior of the RuO$_2$ layers in the itinerant ferromagnetic SrRuO$_3$ and Ru-2122. In both systems the saturation moment at low temperatures should be 2.0 μ$_B$/Ru. However in the case of SrRuO$_3$ and Ru-2122 $M_{sat}$ (at 5 K) are only 1.3 and ~0.9 μ$_B$/Ru respectively. This assumption explains well why the extracted $P_{eff}$=2.15 μ$_B$ in the paramagnetic range is ~ 2.4 larger than $M_{sat}$, and why no magnetic peaks where observed in the neutron diffraction studies. Recently, extensive ac studies provide strong evidence for magnetic frustration in Ru-2122; thus, the possibility of a spin-glass behavior cannot be excluded. This model is to be contrasted with our interpretation of the existence of long-range ordering in our system as discussed above. Further studies such as $^{99}$Ru Mossbauer spectroscopy studies are warranted to investigate the magnetic structure of the Ru-2122 system.

**Acknowledgments** This research was supported by the Israel Academy of Science and Technology and by the Klachky Foundation for Superconductivity.

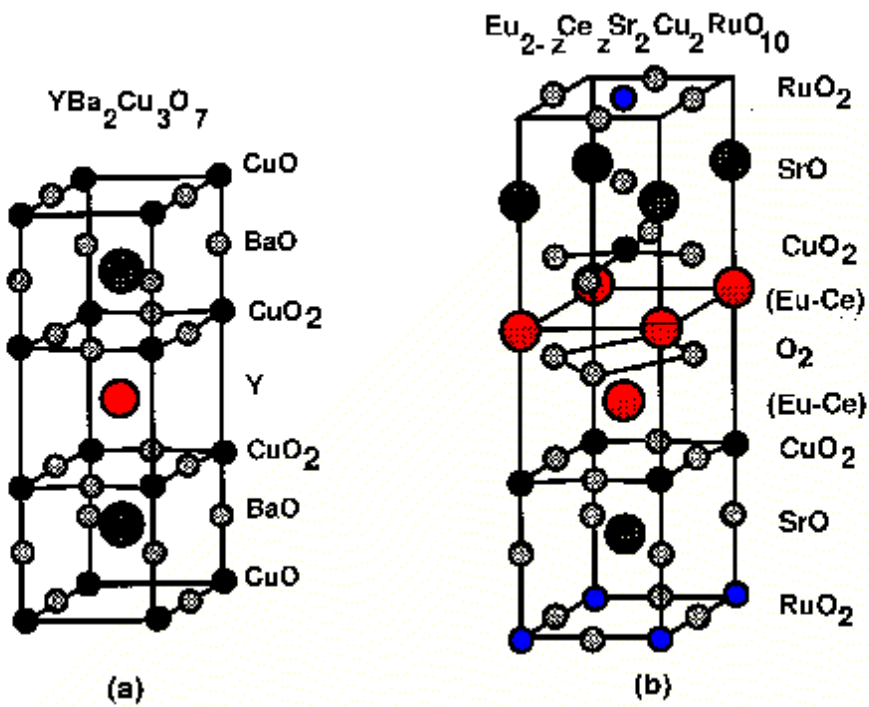

**Fig. 1.** The crystal structure of $YBa_2Cu_3O_7$ and $Eu_{1.5}Ce_{0.5}RuSr_2Cu_2O_{10-\delta}$

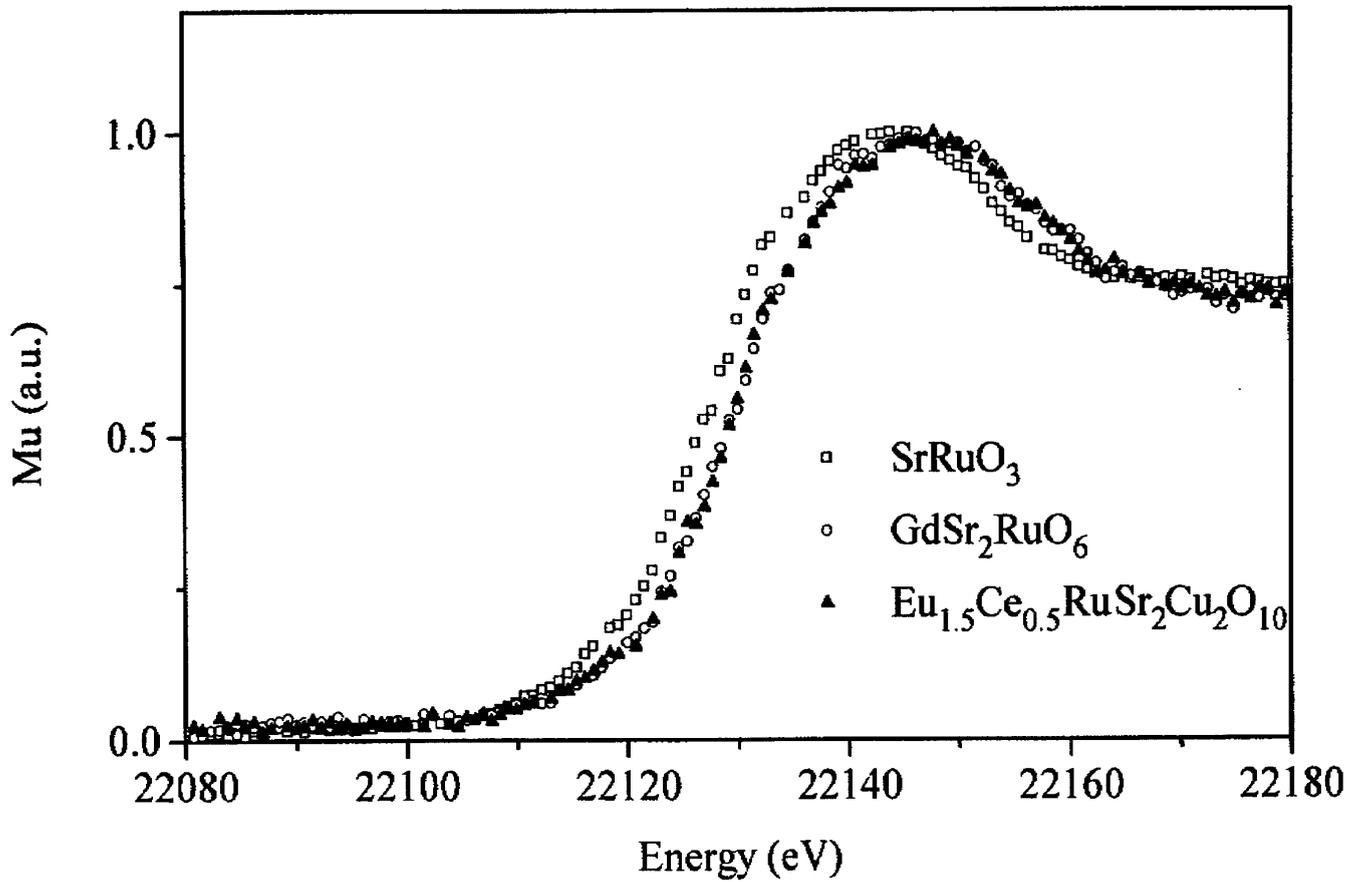

**Fig. 2.** XAS spectra at the K edge of Ru of Ru-2122 and reference compounds.

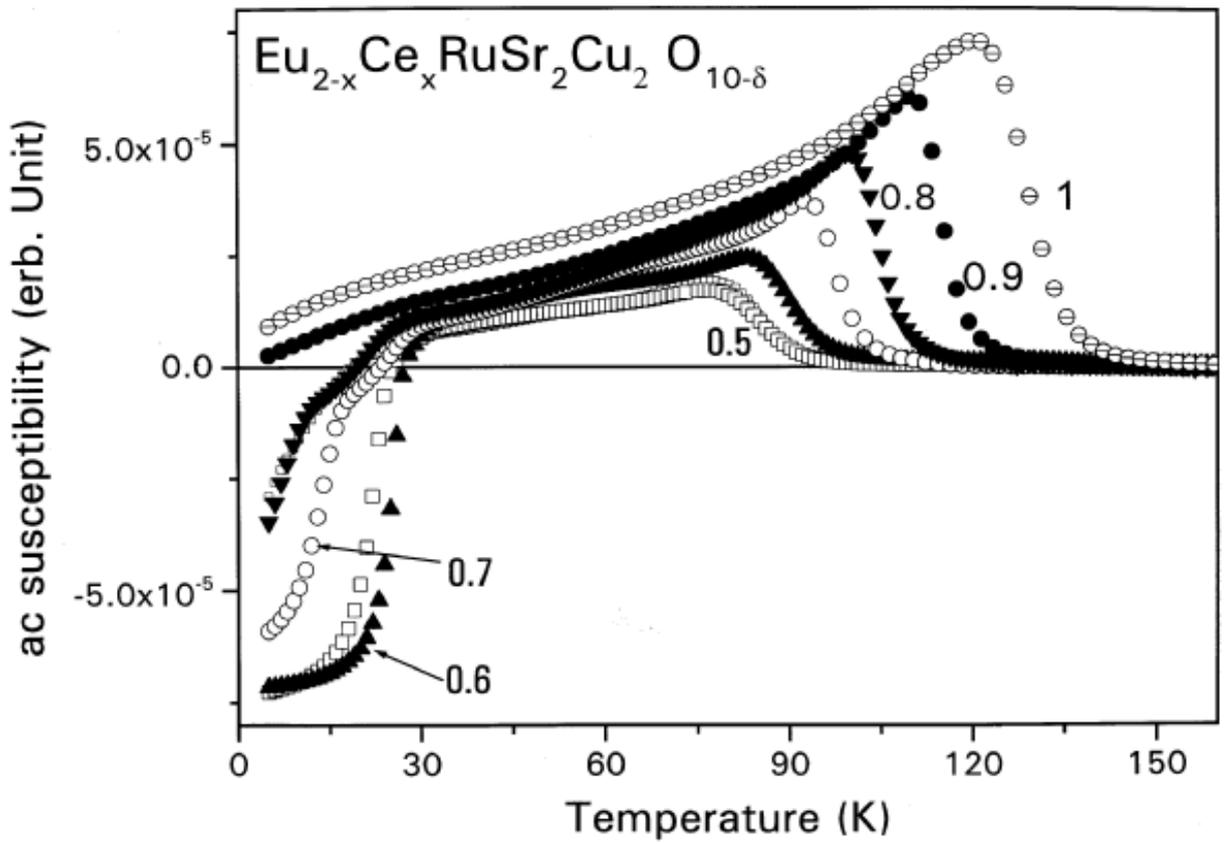

**Fig.3**. Normalized ac susceptibility, ($H_{dc}$ =0) of asp $Eu_{2-x}Ce_xRuSr_2Cu_2O_{10-\delta}$ samples. Note, the absence of SC in the two x=1 and x=0.9 materials.

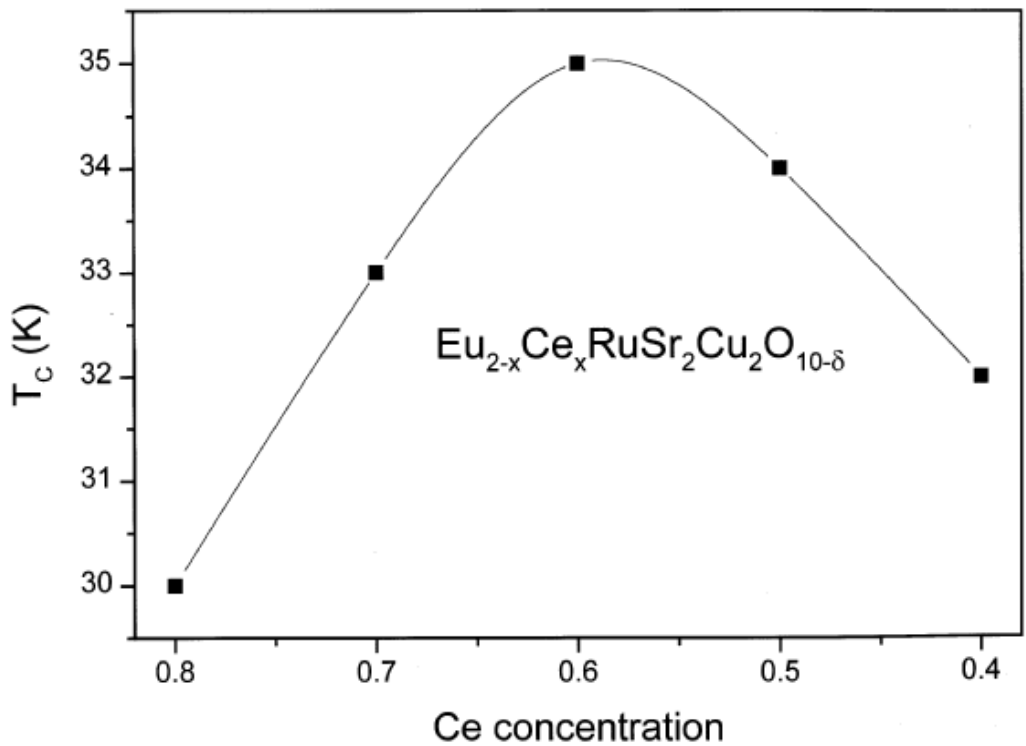

**Fig. 4.** The bell shape SC onset temperature as a function of Ce content.

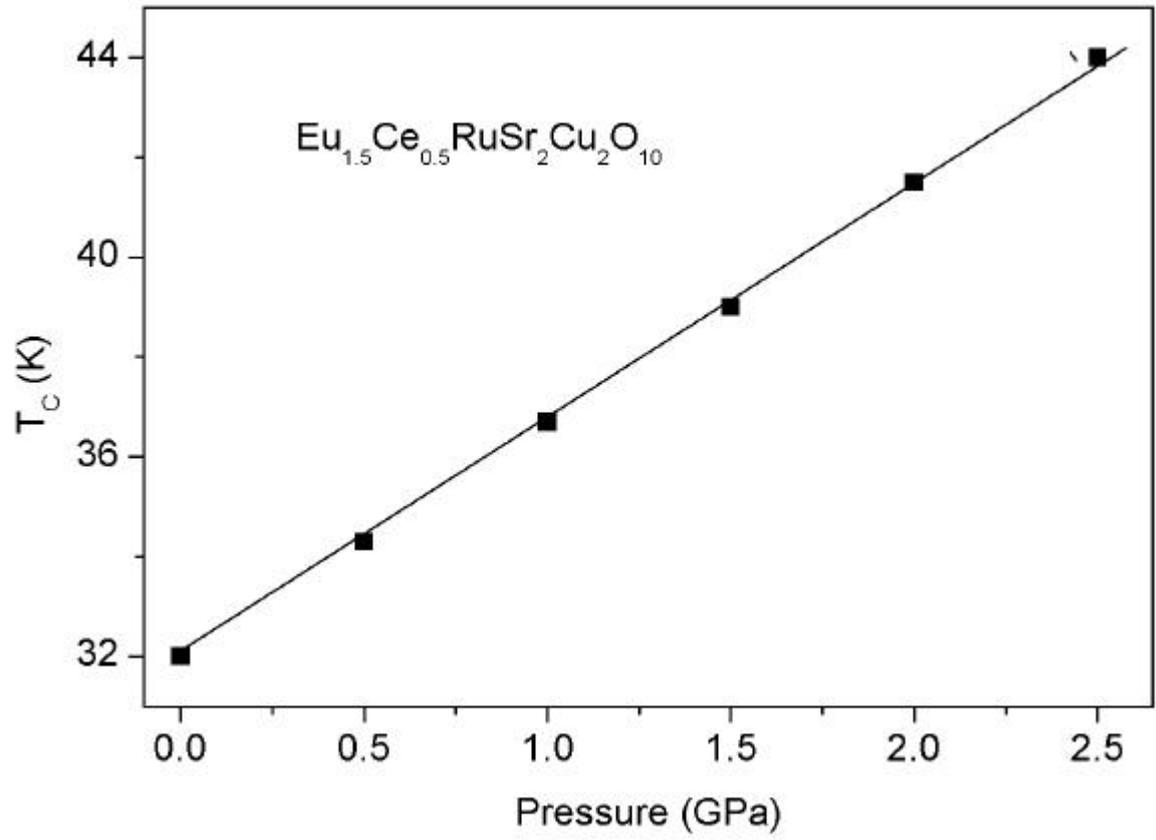

**Fig. 5**. Pressure dependence of $T_C$ of $Eu_{1.5}Ce_{0.5}RuSr_2Cu_2O_{10-\delta}$

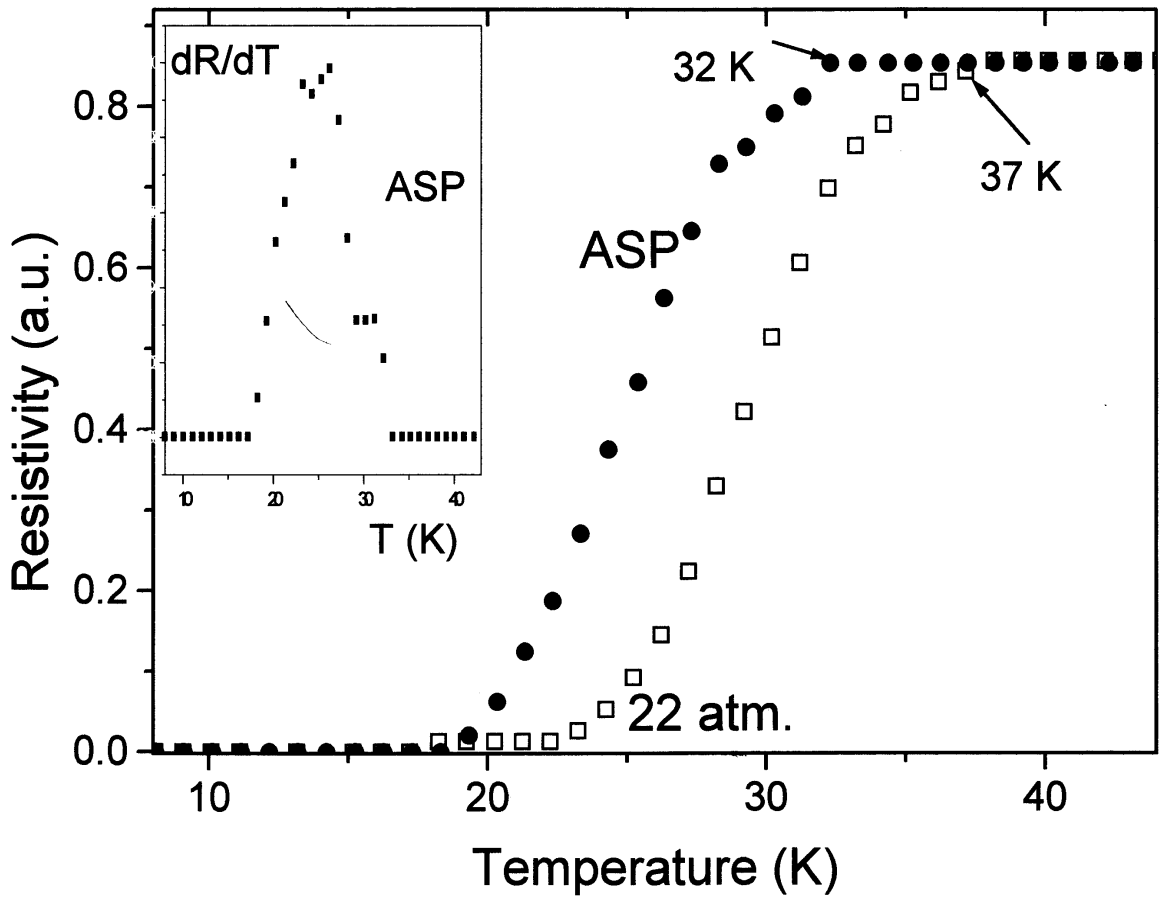

**Fig. 6.** Normalized resistivity measured at H=0 of the asp Ru-2122 and the sample annealed under 22 atm. The inset shows the derivative of the R(T) for the asp sample.

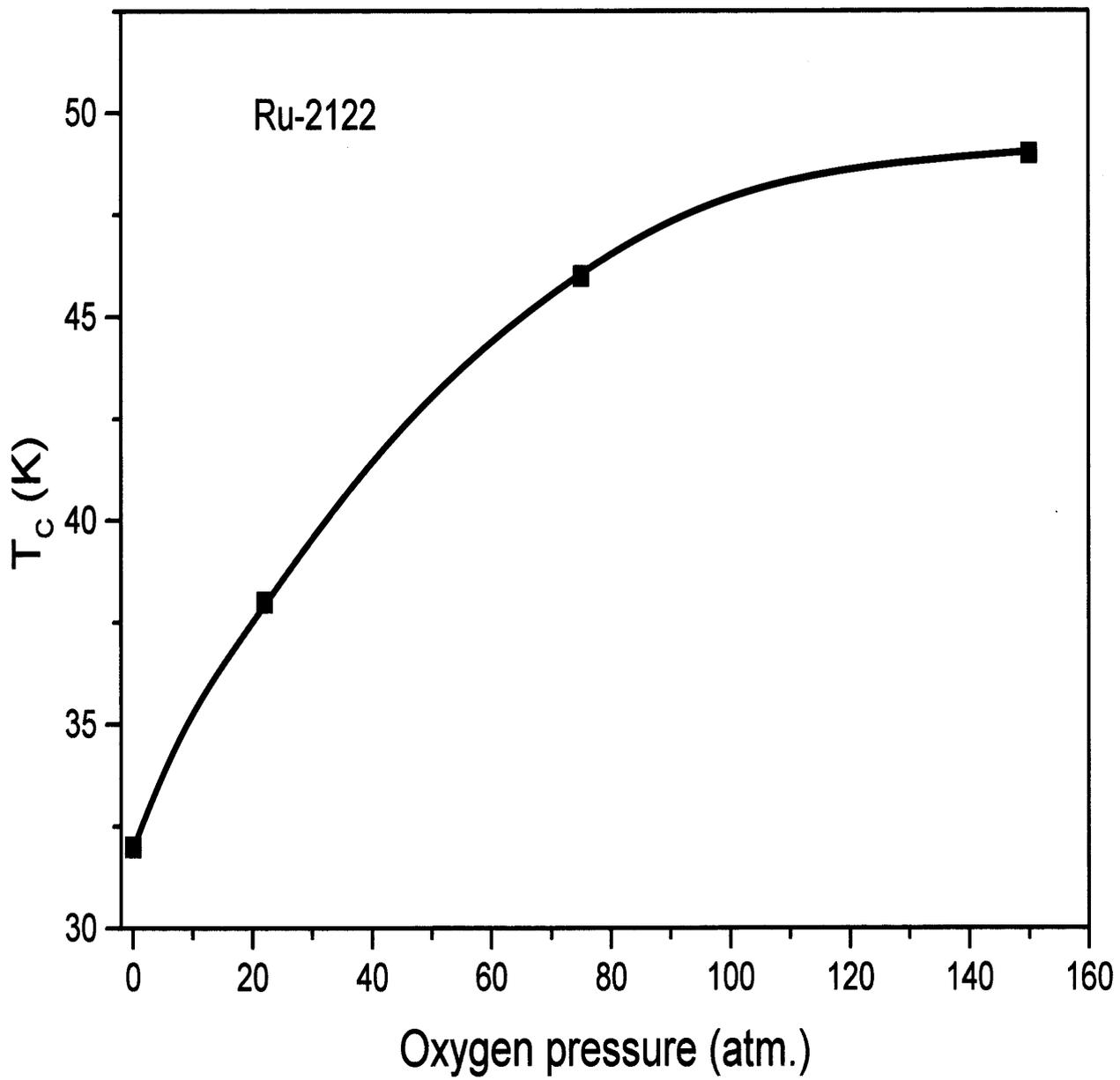

Fig. 7. The effect of the annealing oxygen pressure on $T_C$ of $Eu_{1.5}Ce_{0.5}RuSr_2Cu_2O_{10-\delta}$

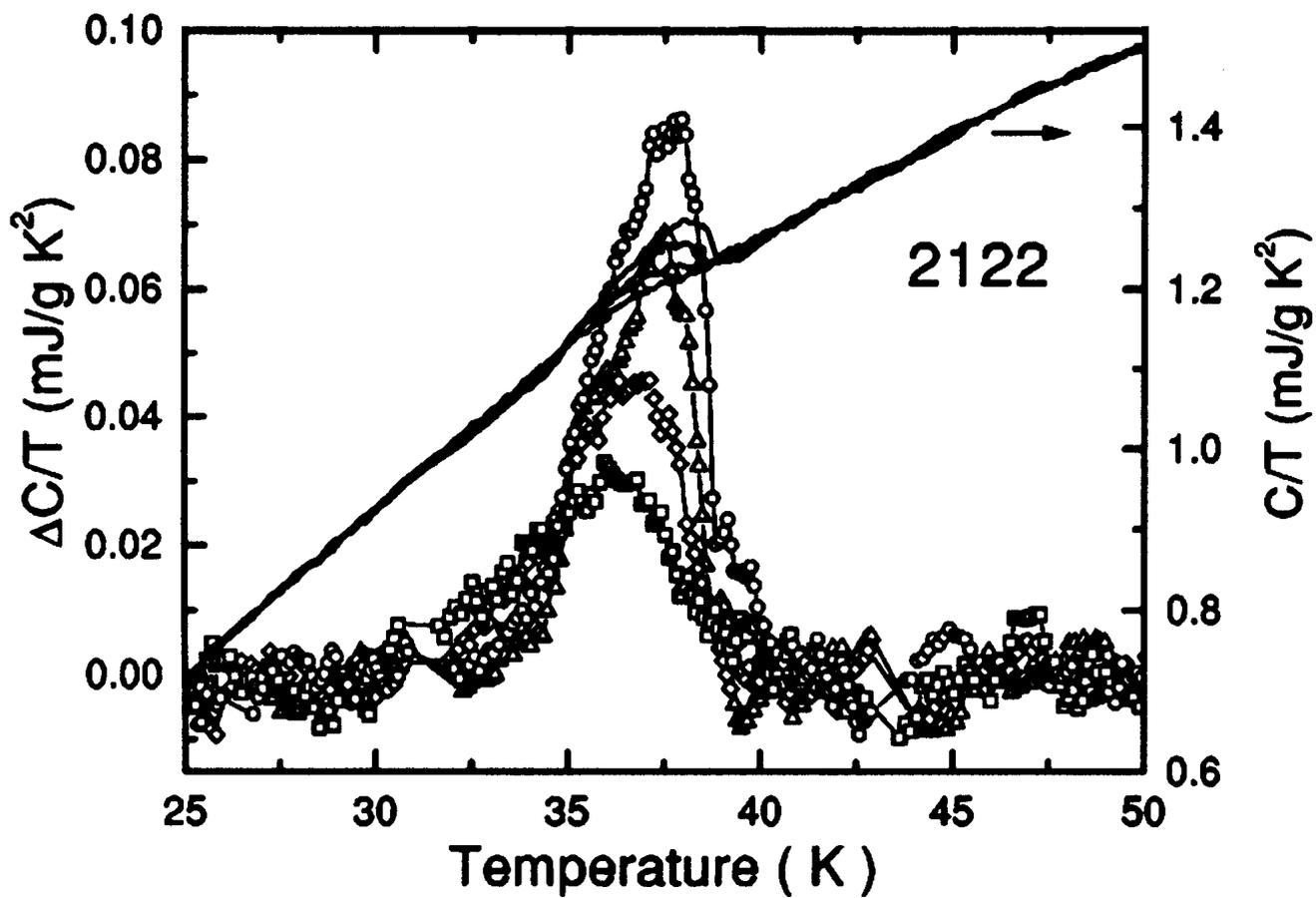

Fig. 8. The specific heat C/T and $\Delta C_P/T$ in magnetic fields up to 6T of $Gd_{1.4}Ce_{0.6}RuSr_2Cu_2O_{10-\delta}$  See Ref. [6]

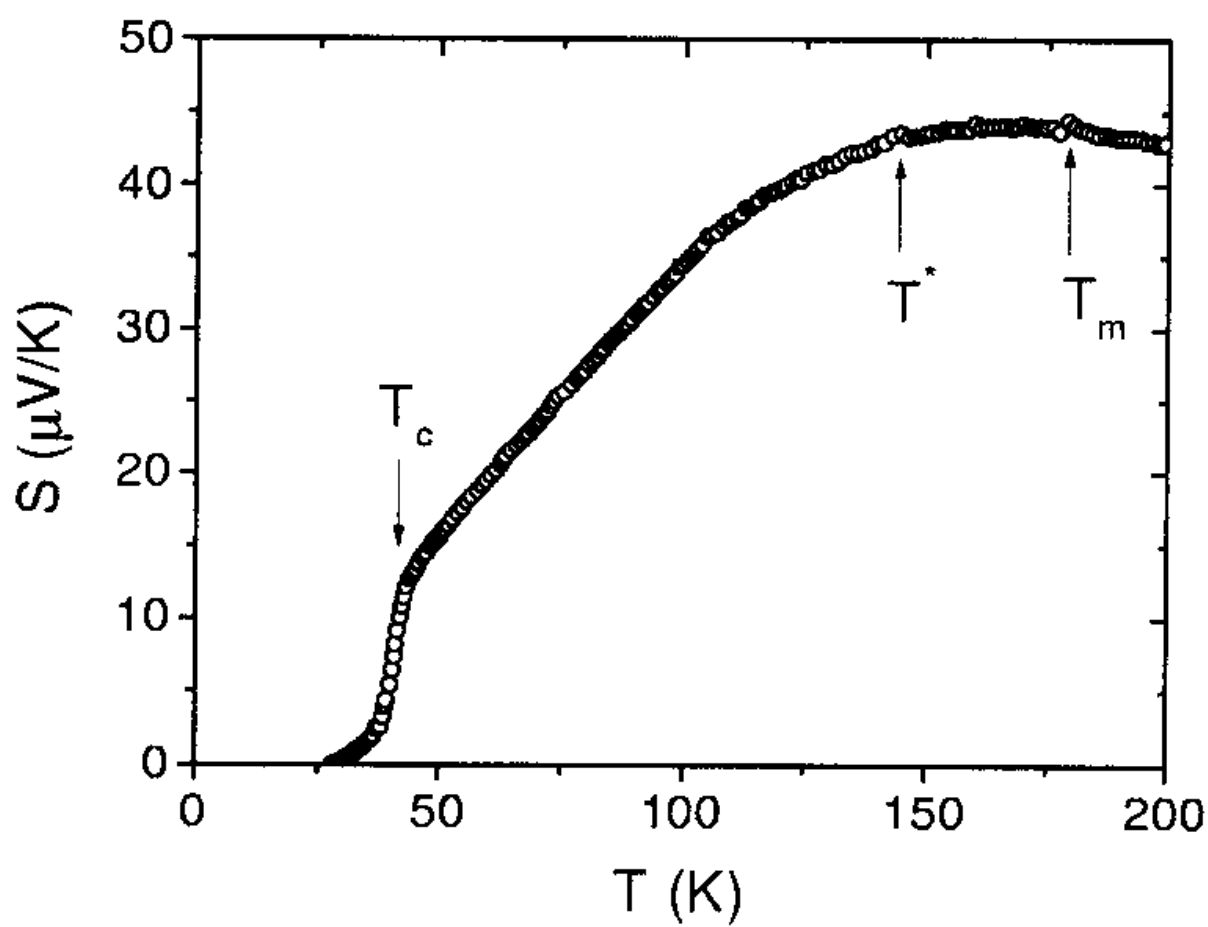

**Fig. 9** Therm-power S versus temperature of $Eu_{1.5}Ce_{0.5}RuSr_2Cu_2O_{10-\delta}$

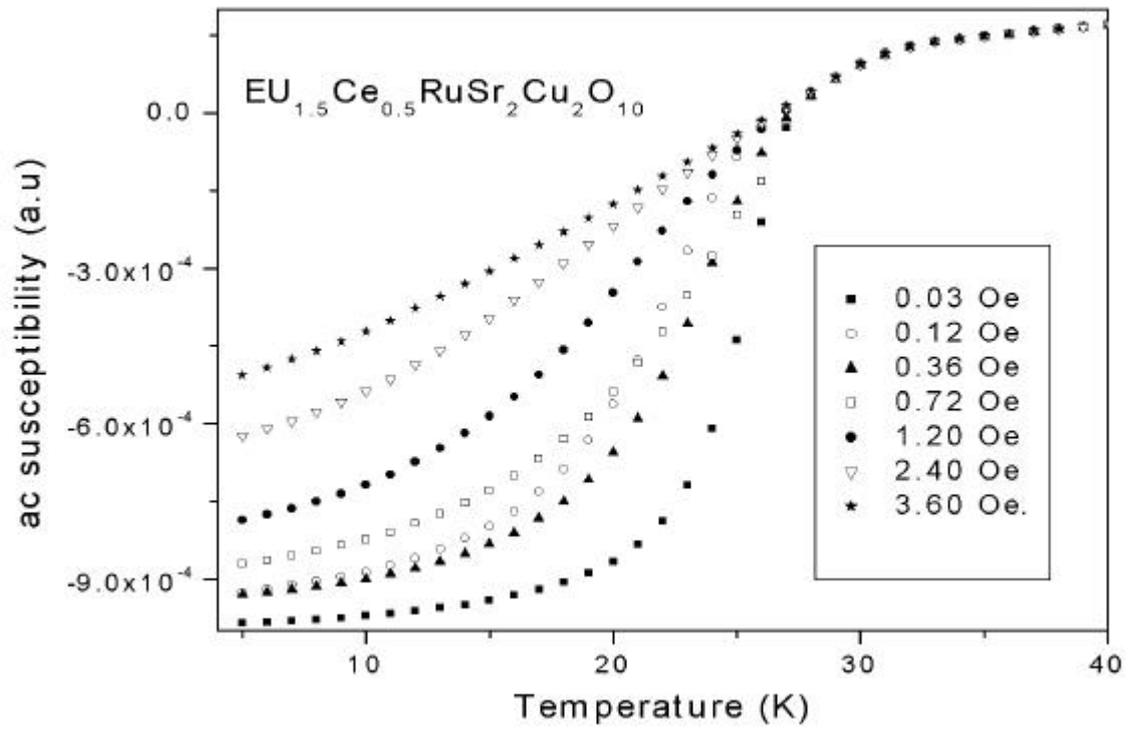

**Fig. 10** Real ac susceptibility curves at various amplitudes ($H_{ac}$) of $Eu_{1.5}Ce_{0.5}RuSr_2Cu_2O_{10-\delta}$

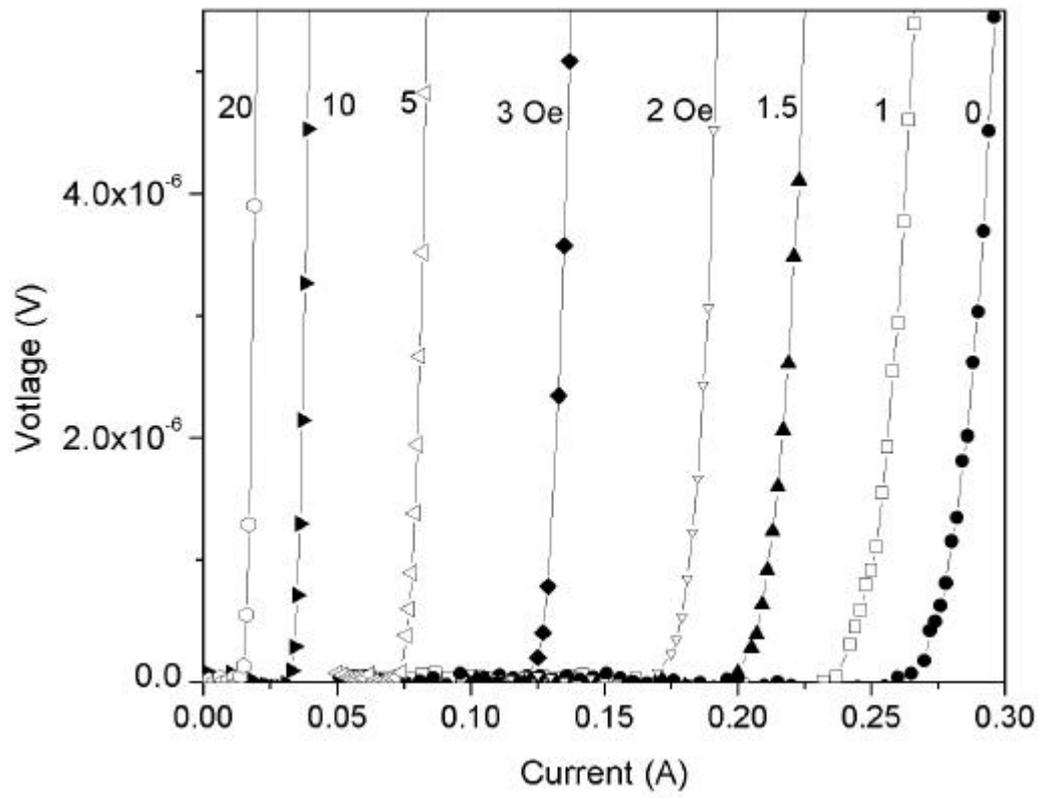

**Fig.11**. Current Voltage curves at 5 K for annealed $RuSr_2Gd_{1.4}Ce_{0.6}Cu_2O_{10}$ at various applied fields in Oe (marked in the figure).

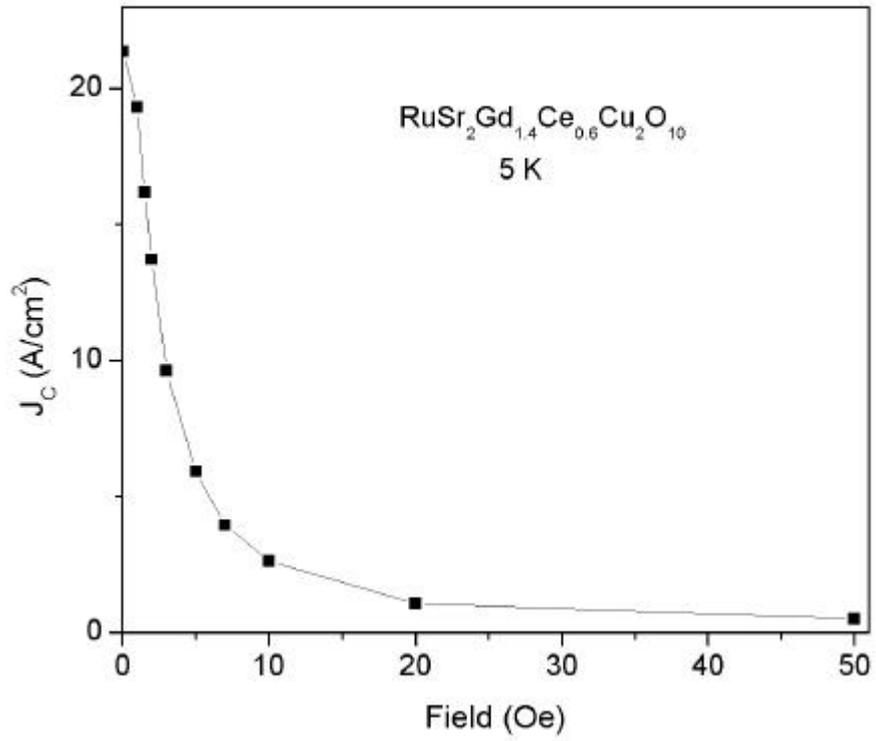

**Fig. 12** The field dependence of the critical current density at 5 K.

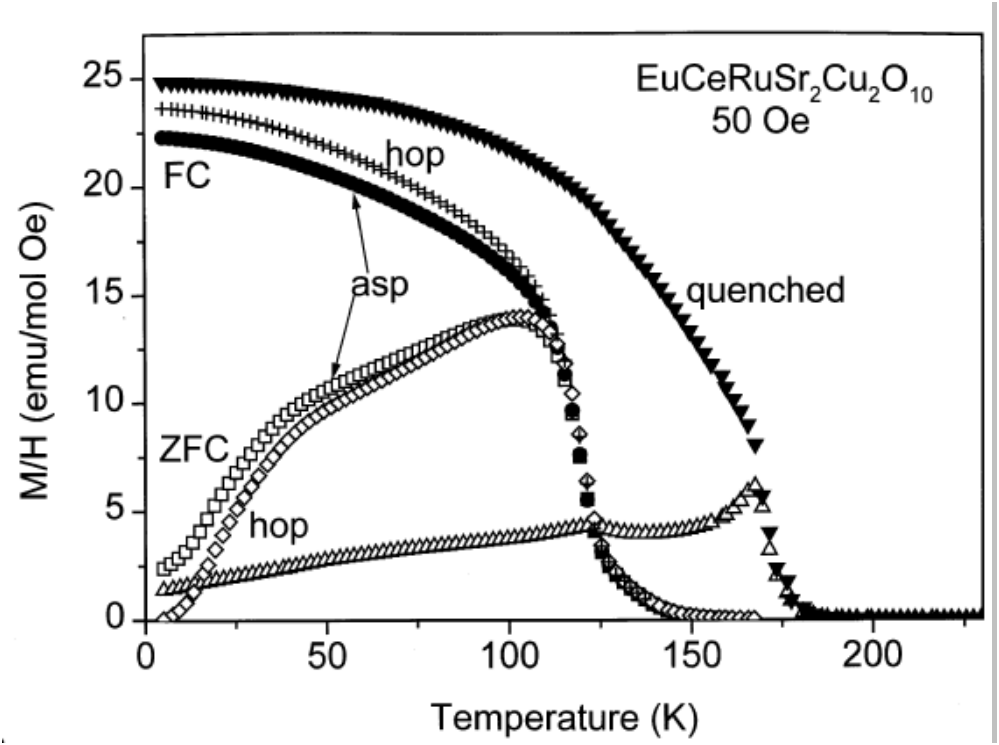

**Fig. 13**. ZFC and FC susceptibility curves for EuCeRuSr$_2$Cu$_2$O$_{10}$ samples,

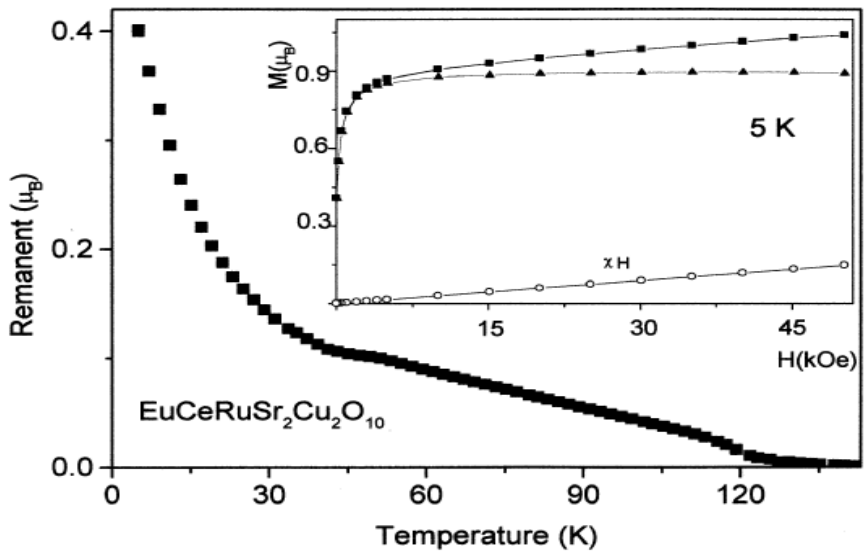

**Fig. 14**. The temperature dependence of 5 K remanent moment of $EuCeRuSr_2Cu_2O_{10}$
The inset shows the high field magnetization and the saturation moment at 5 K.

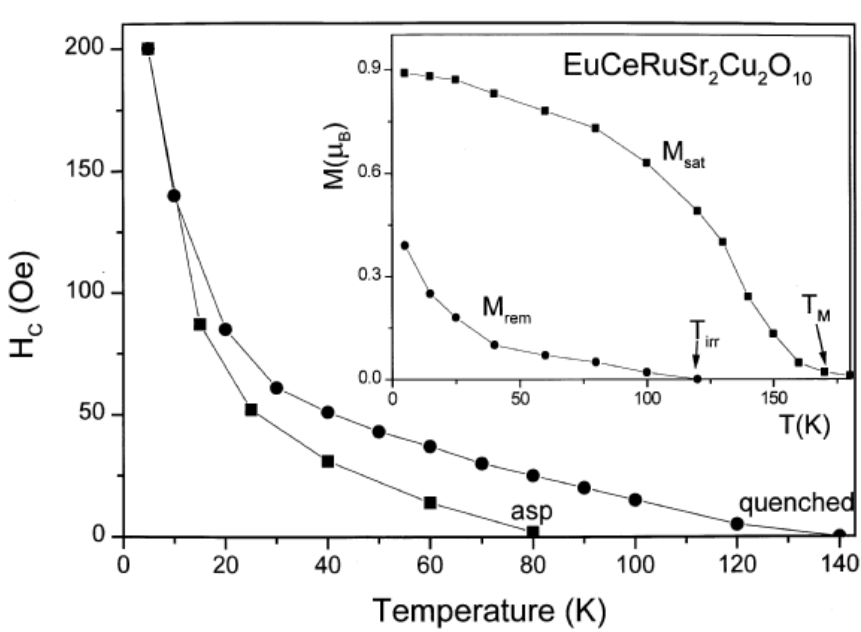

**Fig. 15**. The coercive field $H_C$ as function of temperature for asp and quenched
$EuCeRuSr_2Cu_2O_{10}$ samples. The inset shows the temperature dependence of the
saturation and the remanent moments.

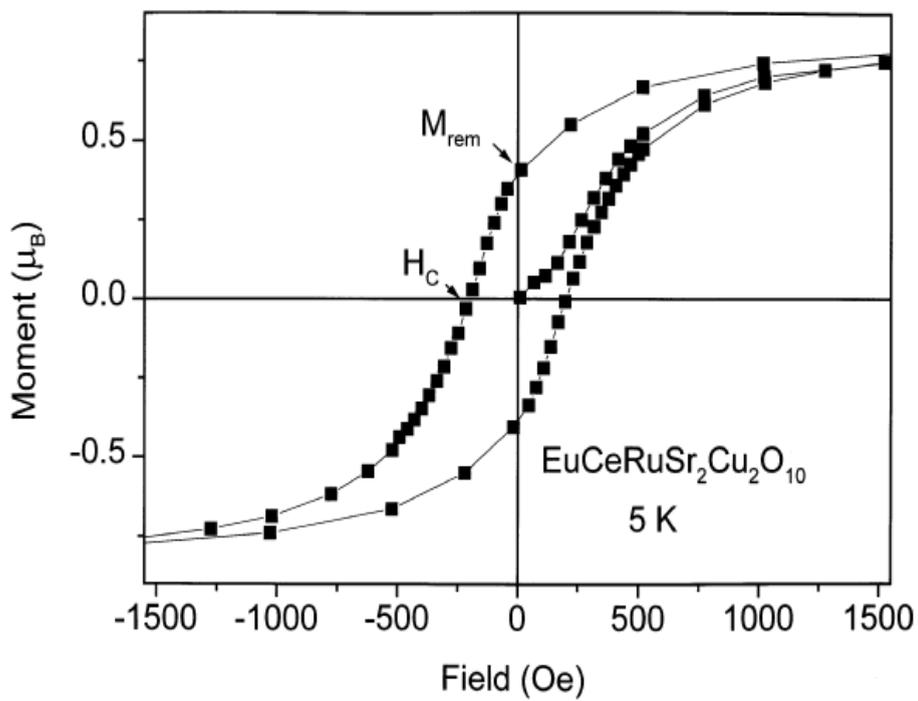

**Fig. 16**. The hysteresis low field loop at 5 K.

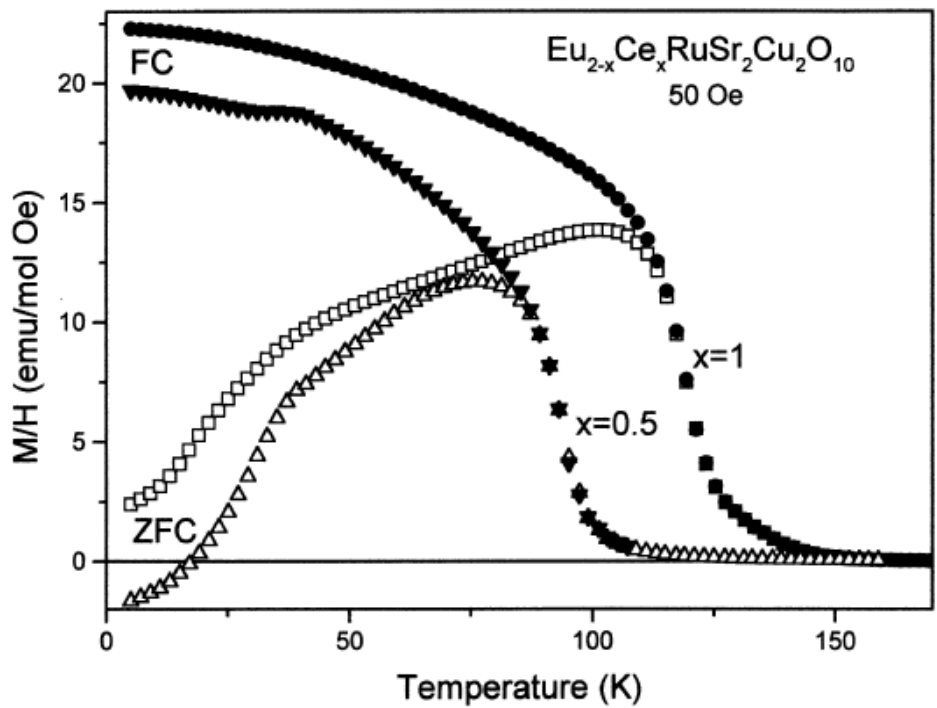

**Fig. 17**. ZFC and FC susceptibility curves for asp x=1 and x=0.5 samples.

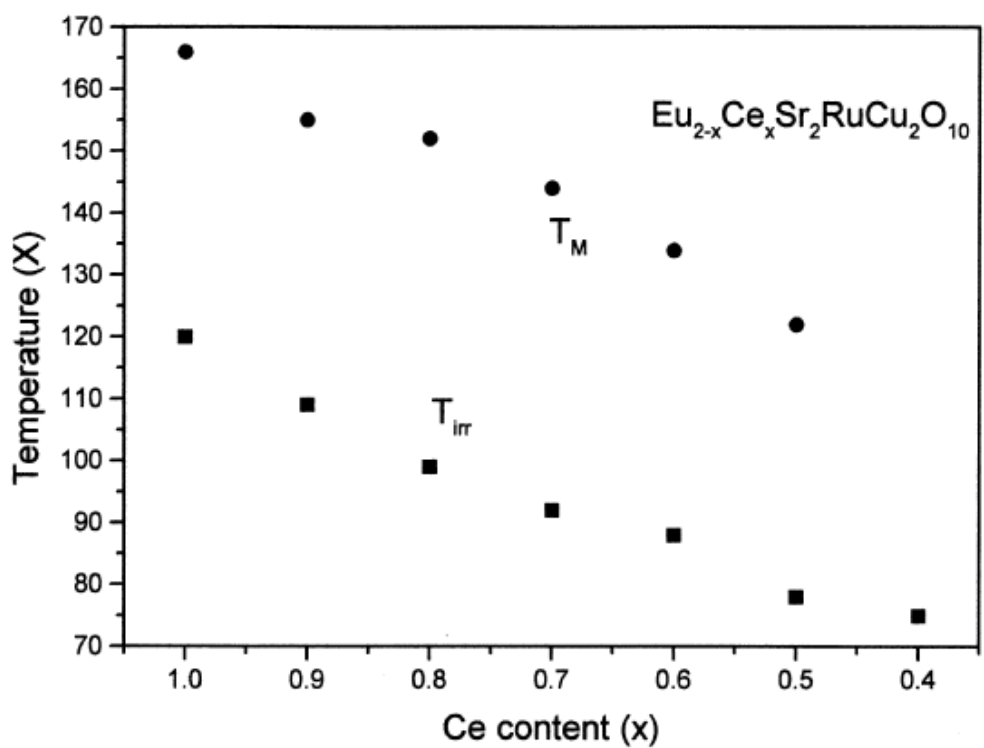

**Fig. 18.** $T_M$ and $T_{irr}$ as function of Ce in $Eu_{2-x}Ce_xRuSr_2Cu_2O_{10}$

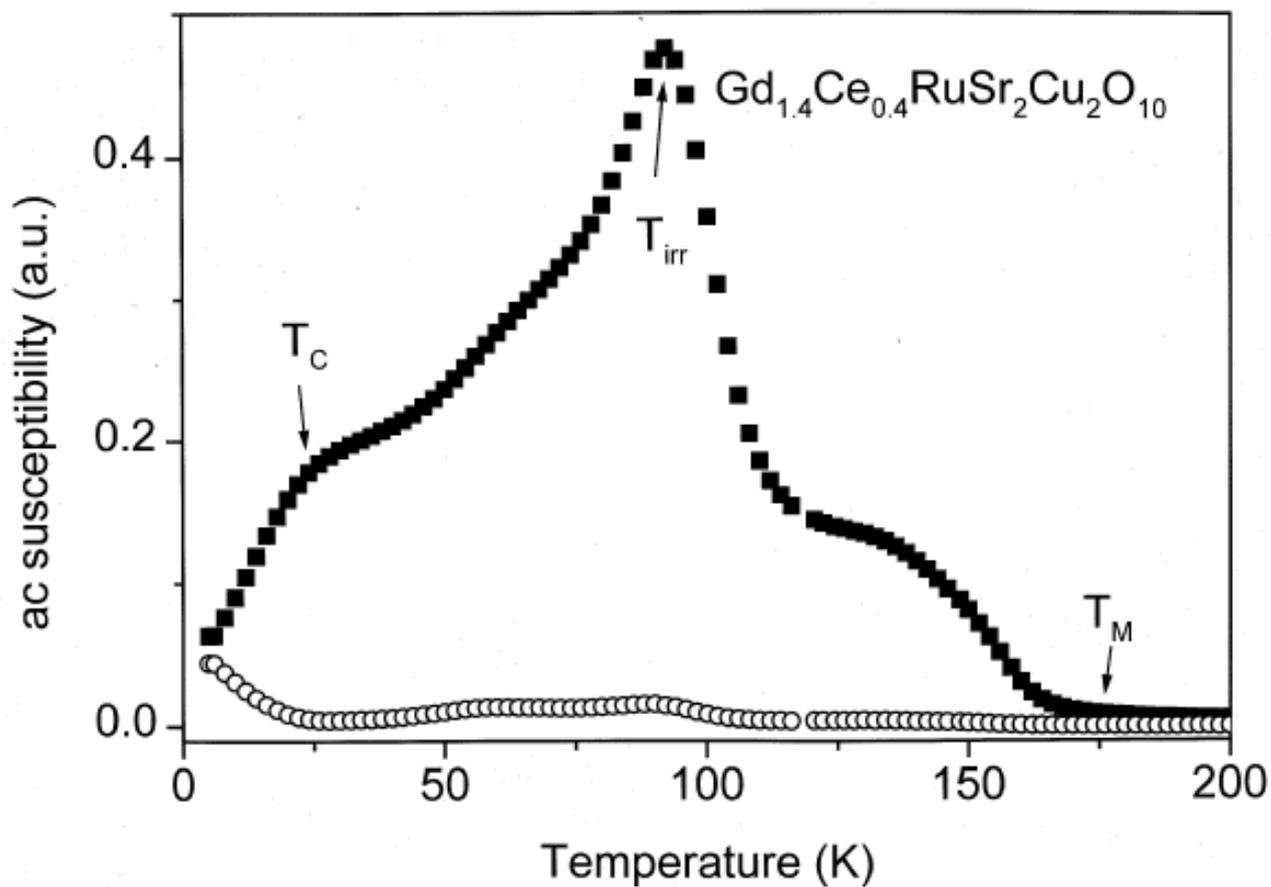

**Fig. 19.** Real and imaginary ac susceptibility of $Gd_{1.4}Ce_{0.6}RuSr_2Cu_2O_{10-\delta}$

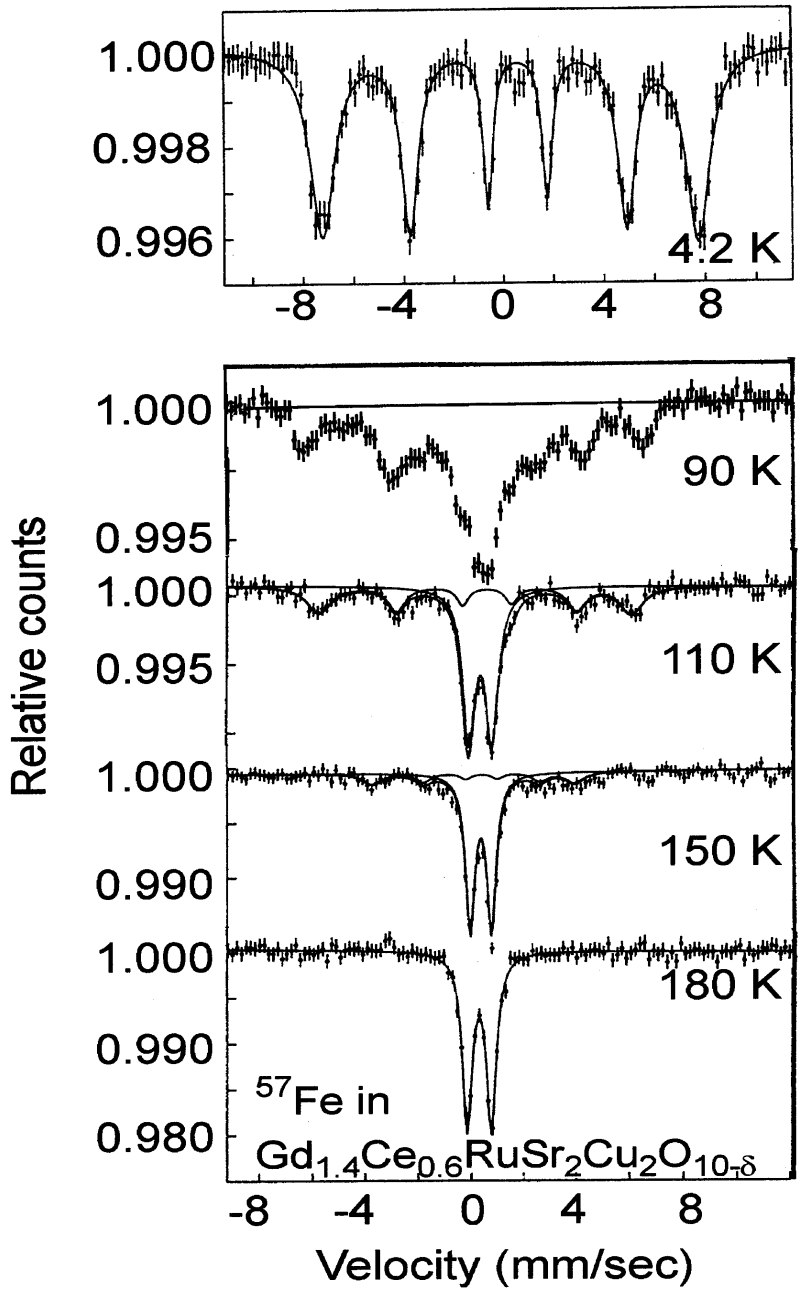

**Fig. 20.** Mossbauer spectra of Fe doped in $Gd_{1.4}Ce_{0.6}RuSr_2Cu_2O_{10-\delta}$.

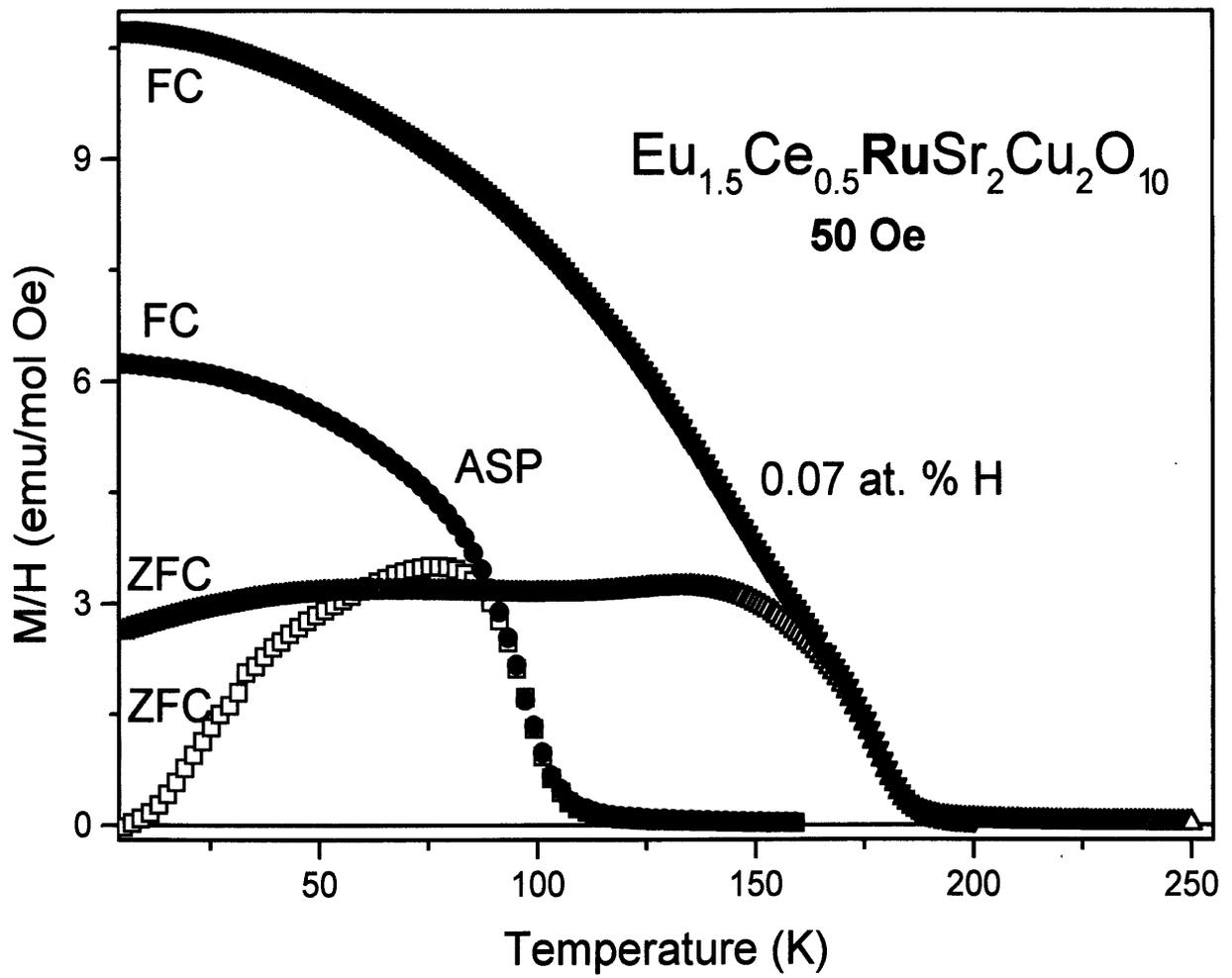

**Fig.21..** ZFC and FC susceptibility curves for asp and Ru-2122H$_{0.07}$.

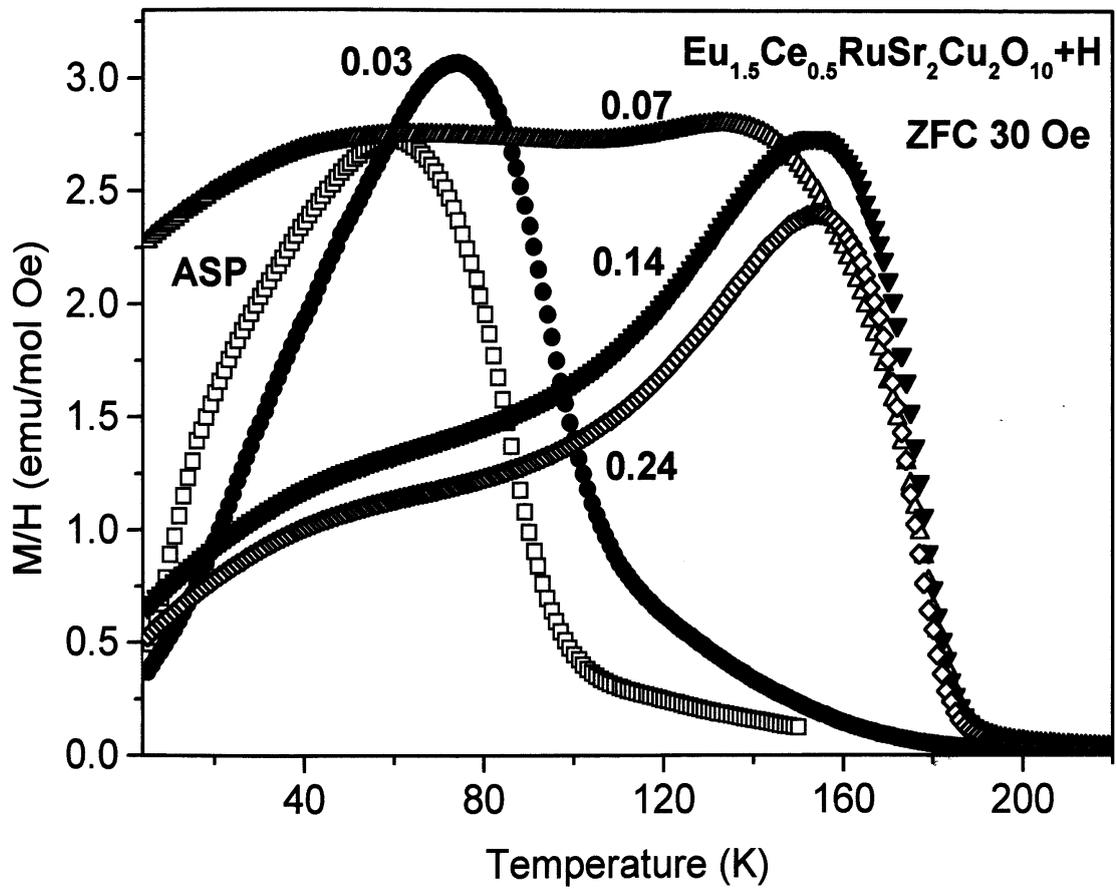

Fig. 22. ZFC susceptibility curves for various hydrogen loaded samples.